\pdfoutput=1
\documentclass[runningheads]{llncs}
\usepackage[T1]{fontenc}
\usepackage{graphicx}
\usepackage{amsmath}
\usepackage{subcaption}
\usepackage{booktabs}
\usepackage{cite}
\usepackage{comment}
\usepackage{longtable}
\usepackage{multirow}
\usepackage{listings}
\begin{document}

\title{MorphoOrgaAgent: A Foundation-Model-Based Multi-Agent System for Autonomous Organoid Analysis}
\titlerunning{MorphoOrgaAgent}
\newif\ifanonymous
%\anonymoustrue

\ifanonymous
  \author{Anonymous Author(s)}
  \authorrunning{Anonymous Author(s)}
  \institute{Anonymous Institution(s)}
\else
\author{Hanyi Zhang\inst{1,3\dagger} \and 
Maximilian Hoermann\inst{1,2\dagger} \and 
Lion J. Gleiter\inst{1,2} \and 
Yiling Xu\inst{3} \and
Bettina Katalin Budai\inst{3} \and
Hans-Ulrich Kauczor\inst{3} \and
Carsten Marr\inst{1,4,5,6,7} \and
Tingying Peng\inst{1,2*}
}
\index{Zhang, Hanyi}
\index{Hoermann, Maximilian}
\index{Gleiter, Lion J.}
\index{Xu, Yiling}
\index{Budai, Bettina Katalin}
\index{Kauczor, Hans-Ulrich}
\index{Marr, Carsten}
\index{Peng, Tingying}
\authorrunning{H. Zhang and M. Hoermann}

\institute{Helmholtz AI, Helmholtz Munich - German Research Center for Environmental Health, Neuherberg, Germany \and
School of Computation, Information and Technology, Technical University of Munich, Munich, Germany \and
Department of Diagnostic and Interventional Radiology, University Hospital Heidelberg, Heidelberg, Germany \and
Institute of AI for Health, Helmholtz Munich - German Research Center for Environmental Health, Neuherberg, Germany \and
Department of Medicine III, Ludwig-Maximilian-University Hospital, Munich, Germany \and
Department of Physics, Ludwig-Maximilian-University, Munich, Germany \and
DKTK, German Cancer Consortium, Heidelberg, Germany
\\
\email{tingying.peng@helmholtz-munich.de}}

\fi
\maketitle

\begingroup
\renewcommand{\thefootnote}{}
\footnotetext{$^{\dagger}$ Equal contribution.}
\footnotetext{$^{*}$ Corresponding author.}
\footnotetext{Preprint. Accepted at the 2nd Agentic AI for Medicine Workshop at MICCAI 2026.}
\endgroup
\begin{abstract}
Organoids are three-dimensional tissue models whose morphology provides important insights into tumor development, disease progression, and drug testing. Extracting these morphological features relies heavily on manual segmentation, which is time-consuming and labor-intensive. Furthermore, performing quantitative statistical analysis typically requires custom coding skills and a mathematical background, presenting a major barrier for experimental biologists. 
To address these challenges, we introduce MorphoOrgaAgent, a multi-agent framework that achieves zero-shot organoid segmentation, automated data analysis, and report generation based on natural language input. The framework consists mainly of three core components: a TaskUnderstandingAgent that identifies requested measurements and visualization types; a hybrid segmentation module that combines Cellpose-derived geometric prompts with text prompts to guide SAM3 for zero-shot organoid instance segmentation; and a ReportAgent that computes quantitative metrics and compiles them alongside generated visualizations into a structured report. We further introduce MorphoOrgaVQA, a benchmark designed for quantitative evaluation of agent systems in organoid morphology analysis. Experimental results demonstrate that MorphoOrgaAgent handles both explicit and descriptive user requests, produces measurements closely matching ground truth, and generates complete analysis reports without requiring manual programming. The complete source code and MorphoOrgaVQA benchmark are publicly available at \url{https://github.com/peng-lab/MorphoOrgaAgent}.
\end{abstract}

\section{Introduction}
Organoids are three-dimensional, self-organizing tissue-like structures that capture key aspects of the structure and function of human or animal organs. Their unique properties make them promising platforms for clinical diagnostics, personalized medicine, disease modeling, and high-throughput drug screening~\cite{ref_organoids}. Morphological features such as organoid size and shape, together with organoid count, provide informative quantitative readouts of growth, culture state, and responses to experimental perturbations.~\cite{ref_orgaquant,ref_tellu,ref_orgaextractor,ref_orgasegment,ref_organoseg2}.
Precise instance segmentation serves as the cornerstone for extracting such biologically meaningful morphological measurements from microscopy images. To address this need, various specialized deep learning and conventional image-processing methods have been developed. For instance, models such as OrganoID~\cite{ref_organoid_model}, OrganoSeg~\cite{ref_organoseg,ref_organoseg2}, OrgaQuant~\cite{ref_orgaquant}, OrgaExtractor~\cite{ref_orgaextractor}, OrgaSegment~\cite{ref_orgasegment}, Tellu~\cite{ref_tellu} and TransOrga-plus~\cite{ref_transorgaplus}
provide automated detection, instance segmentation, or morphological classification for specific organoid types, while tools like NOA~\cite{ref_noa_isbi} offer graphical user interfaces to facilitate analysis. 
Thus, these methods have enabled the systematic analysis of large organoid imaging datasets by replacing labor-intensive manual measurements with scalable and reproducible image analysis pipelines. Despite these valuable contributions, most existing pipelines heavily rely on task-specific or dataset-specific training, limiting their zero-shot generalization capabilities across diverse organoid phenotypes and varied imaging conditions.
\\
Recently, large language model (LLM)-based agents have emerged as a powerful paradigm to automate complex workflows by translating natural-language requests into executable tasks. In the biomedical domain, impressive agentic frameworks have been introduced for specialized applications: BioMedAgent~\cite{ref_biomedagent} chains diverse bioinformatics tools to solve data-driven tasks, CellAgent~\cite{ref_cellagent} automates single-cell data analysis, and the BioImage.IO Chatbot~\cite{lei2024bioimageio} leverages multi-agent assistance to orchestrate bioimage analysis tools. For image-processing workflows, Agentic-J~\cite{ref_agenticj} integrates LLM reasoning with ImageJ within containerized environments, while Omega~\cite{royer2024omega} provides a Napari-based agent interface for interactive image analysis. In computational pathology, innovative frameworks such as SPARK~\cite{ref_spark} autonomously code and validate biomarker concepts without model retraining, and PathAgent~\cite{ref_pathagent} delivers transparent whole-slide image analysis through explicit reasoning traces. Furthermore, systems like Agentic Lab~\cite{ref_agenticlab} demonstrate the utility of LLMs in coordinating protocol design, laboratory guidance, and organoid phenotyping. Despite these remarkable advancements, a multi-agent framework capable of achieving zero-shot organoid segmentation, automated quantitative analysis, and comprehensive report generation remains lacking.
\\
We therefore introduce MorphoOrgaAgent, an autonomous multi-agent system that combines foundation-model-based zero-shot segmentation with LLM-driven quantitative morphology analysis and report generation. Its workflow consists of three coordinated components: First, the TaskUnderstandingAgent translates the biologist's natural-language request into a structured analysis plan by selecting required measurements and visualizations from a predefined analytical pool, while also formulating tailored instructions to guide the downstream ReportAgent. Next, the Hybrid-Prompt Segmentation Module combines Cellpose-derived geometric prompts with text prompts to guide SAM3 for zero-shot organoid instance segmentation, eliminating the need for tedious manual annotation. Finally, the ReportAgent automatically computes the requested instance- and population-level metrics, generates corresponding visualizations, and compiles all findings into a structured analysis report.
\\
We evaluate MorphoOrgaAgent through both quantitative benchmarks and qualitative comparative analyses. To quantitatively assess performance, we introduce MorphoOrgaVQA, a Visual Question Answering benchmark compiled from three public organoid datasets, featuring morphology-focused queries that address key biological questions. To ensure objective and reproducible evaluation, we release an automated pipeline that generates deterministic ground-truth (GT) answers directly from expert-annotated masks. Furthermore, we conduct qualitative comparisons against existing bioimage frameworks, including Omega~\cite{royer2024omega} and Agentic-J~\cite{ref_agenticj}, on complex analytical tasks. Experimental results demonstrate that MorphoOrgaAgent excels in zero-shot instance segmentation, produces accurate statistical computations, and synthesizes clear, structured, and professional analysis reports. By automating these critical steps, our framework significantly reduces manual annotation effort, eliminates coding requirements, and enables batch analysis for large-scale organoid studies.

\section{Methodology}
\paragraph{System Overview: MorphoOrgaState and Multi-Agent Orchestration}
At the core of our pipeline is MorphoOrgaState, a centralized, JSON-serializable state container that acts as the single source of truth shared across all agents and modules. A new MorphoOrgaState instance is initialized for every user query, storing the input image path and the natural language query at the outset. As execution proceeds through the pipeline, each subagent and module reads from and writes back to this shared state, and the updated state is persisted to disk after the final step, yielding a complete and auditable trace of all intermediate results.
Concretely, the TaskUnderstandingAgent first parses the user's analysis intent and writes the resulting analysis plan, including the target objects, required metrics, and required visualizations, into MorphoOrgaState. The segmentation module is then invoked, and the resulting organoid masks are likewise stored in the state. Quantitative metrics computed from these masks are appended in the same manner. Finally, the ReportAgent consolidates all information accumulated in MorphoOrgaState, including the original and translated query, segmentation results, and computed metrics, into a comprehensive natural language report that directly answers the user's question. To support full reproducibility, the complete system prompts of both LLM-driven subagents are reported verbatim in Appendix~\ref{app:prompts} and released in our GitHub repository.

\begin{figure}[t]
    \centering
    \includegraphics[width=\textwidth]{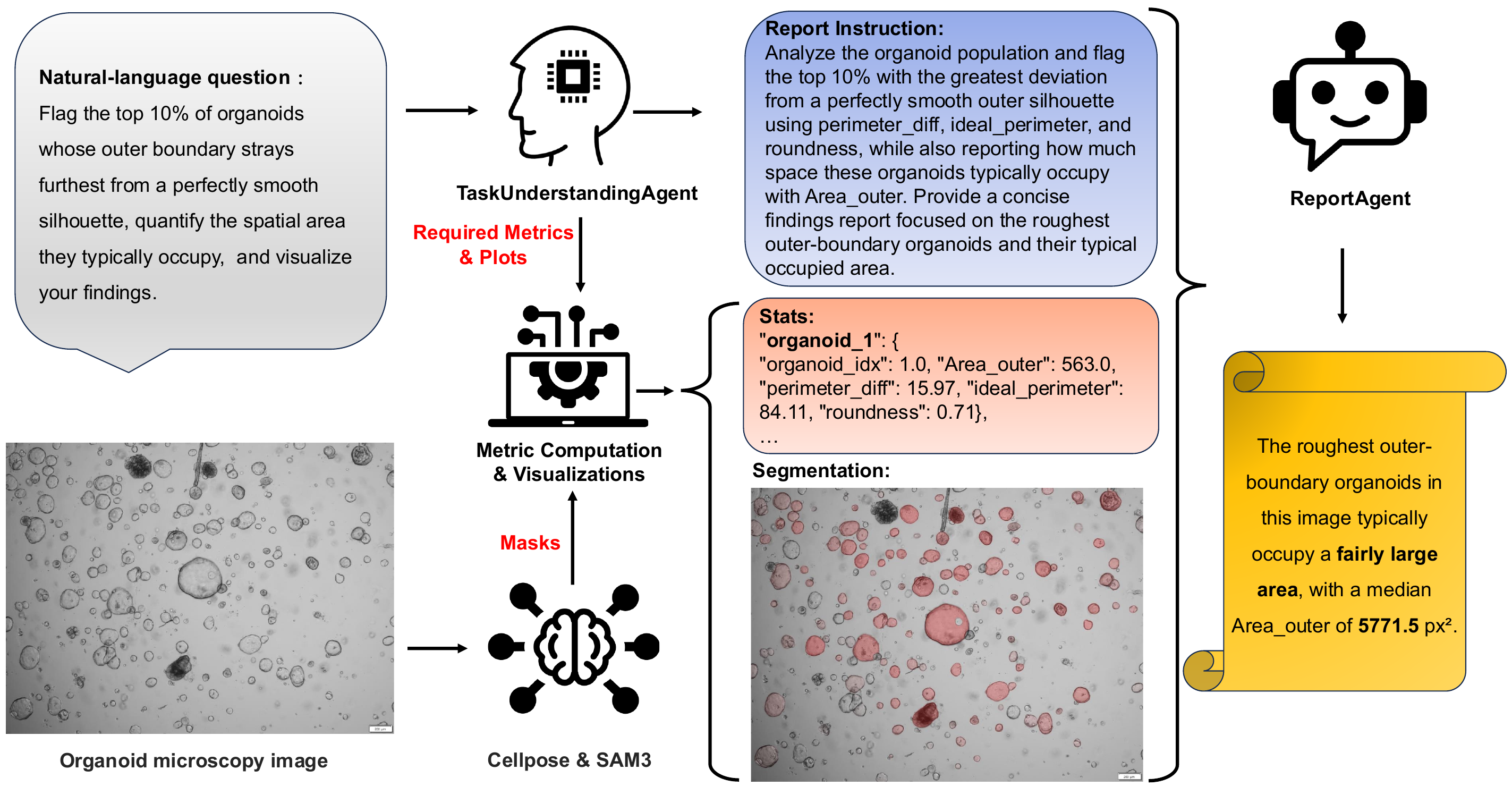}
    \caption{\textbf{Overview of the MorphoOrgaAgent pipeline.} Given a microscopy image and a natural-language query, the TaskUnderstandingAgent translates the request into a structured analysis specification and a report instruction for downstream processing. Cellpose and SAM3 then perform zero-shot instance segmentation. Using the resulting masks and analysis specification, the framework computes the requested metrics and visualizations. Finally, the ReportAgent integrates the original query, segmentation outputs, and quantitative results into a concise, evidence-grounded report.}
    \label{fig:mainfigure}
\end{figure}
\paragraph{TaskUnderstandingAgent}
The TaskUnderstandingAgent performs three main functions. First, it takes MorphoOrgaState as input and interprets the user's natural language query to determine the underlying analysis intent. Second, based on this intent, it selects the necessary metrics and visualizations from a predefined pool of supported metrics and plot types. Third, since the user's query may be ambiguous, imprecise, or contain typographical errors, the agent reformulates it into a more precise, scientifically phrased prompt, which is likewise saved in MorphoOrgaState and later passed to the ReportAgent after the segmentation module has completed. This agent is powered by GPT-5.4-mini.
\paragraph{Hybrid-Prompt Segmentation \& Metric Computation}
To improve generalization beyond organoid-specific segmentation models, we combine Cellpose~\cite{stringer2021cellpose} and SAM 3~\cite{carion2025sam3} in a hybrid-prompt strategy. Although SAM 3 supports text-prompted segmentation, its predominantly natural-image training limits its representation of the domain-specific biological concept of an “organoid”, making text prompts alone insufficient. Cellpose, by contrast, provides useful coarse localization of biological objects in microscopy images but may miss or imprecisely delineate organoids with unfamiliar morphologies. We therefore use Cellpose-derived masks as geometric prompts, together with the text prompt ``cell cluster'' (a description that more accurately characterizes the biological nature of organoids) 
to guide SAM3 toward a refined segmentation. Once segmentation is complete, the resulting masks are collected, and since all morphological statistics are derived from these masks, the system invokes the predefined metric-computation and visualization functions to compute the corresponding statistics. All results are then saved into MorphoOrgaState, serving as input to the final ReportAgent.
\paragraph{ReportAgent}
The ReportAgent reads all information stored in MorphoOrgaState, including the original image, segmentation results, and computed statistics, and answers the user's query refined by the TaskUnderstandingAgent. It generates a final report that directly answers the query and explicitly cites the supporting evidence drawn from MorphoOrgaState. To mitigate hallucination, the ReportAgent is strictly constrained to base its reasoning and answer solely on the information contained in MorphoOrgaState, rather than on its own prior knowledge. This agent is powered by the more capable GPT-5.4.

\section{Experiments}
\subsection{The MorphoOrgaVQA Benchmark}
\label{question_benchmark}
To quantitatively evaluate MorphoOrgaAgent by simulating biologists' real-world requests for organoid morphological analysis, we construct the MorphoOrgaVQA
benchmark. It comprises 16 questions spanning the most commonly used
morphological metrics, including area, perimeter, roughness, and roundness. These 16
questions are organized into two phrasing modes: \emph{clear} and
\emph{open}. In clear-mode questions, the metric name appears explicitly in
the text, whereas in open-mode questions, the metric is only implied. For
example, ``Identify the specific organoid with the maximum outer area in
this field of view, and report both its total cross-sectional surface area
in pixels and its geometric center coordinates x and y'' is a clear
question, since the target metrics (outer area, coordinates $x$ and $y$)
are stated directly. Its open counterpart, ``Locate the most dominant
organoid in this image and evaluate how much footprint it occupies in
pixel coordinates, alongside its center of mass,'' conveys the same
intent through more colloquial, biologist-style phrasing, requiring the
system to infer the underlying metrics from context rather than from
explicit keywords.
To assess the system's ability to perform the population-level statistical
analyses that are of particular importance to biologists, each metric is
evaluated under two complementary use cases: identifying the extremum
(largest) organoid, and stratifying the population into three tiers (small, medium, large). For
example: ``Perform a stratification of this sample into small, medium, and large tiers based on individual outer area distributions. Provide the mean outer area value calculated for each of the three tiers.''
The full set of benchmark questions is provided in Appendix~\ref{app:benchmark} and released in machine-readable form in our public GitHub repository. We further provide a script that computes GT answers directly from the expert-annotated masks, ensuring an objective and reproducible evaluation protocol. The benchmark is built on test sets from three high-quality, publicly available datasets with expert annotations, namely OrganoID \cite{ref_organoid_model}, OrgaExtractor \cite{ref_orgaextractor}, and OrgaSegment \cite{ref_orgasegment}, totaling 69 images with corresponding masks. Running all 16 questions on every image yields $16 \times 69 = 1{,}104$ question--answer pairs.

\subsection{Quantitative Evaluation of MorphoOrgaAgent}
To evaluate the numerical precision and execution fidelity of MorphoOrgaAgent, we benchmark its predictions against deterministic GT values computed directly from expert-annotated instance masks across the entire benchmark. For each quantitative query, we measure performance using the absolute percentage error (APE) $\epsilon = \left| (V_{\text{pred}} - V_{\text{GT}})/V_{\text{GT}} \right| \times 100\%$ where $V_{\text{pred}}$ denotes the scalar value predicted by the agent, and $V_{\text{GT}}$ denotes the corresponding GT value computed analytically from the annotated mask. As outlined in Section~\ref{question_benchmark}, we evaluate model performance across two distinct query formats: single-extremum queries and tertile-mean estimations. The main quantitative evaluation results across all morphological metrics and query types are summarized in Table~\ref{tab:median_ape_results}. Notably, our multi-agent system achieves exceptional precision on Roundness, yielding a median APE of less than $1.6\%$ for single-extremum queries ($1.54\%$ clear, $1.56\%$ open) and remaining under $2.6\%$ for tertile mean queries ($2.48\%$ clear, $2.51\%$ open). This demonstrates the system's strong zero-shot capability in capturing global morphological roundness and geometric regularity. For fundamental spatial metrics such as Area and Perimeter, the system reliably locates single-extremum targets with median APE ranging between $15.36\%$ and $17.40\%$. While Area predictions exhibit a larger shift when aggregating subpopulation statistics (tertile mean APE of $38.74\%$ for clear prompts), Perimeter maintains stronger stability across aggregated cohorts, recording a median APE of approximately $22.0\%$. Conversely, Roughness poses the most significant technical challenge, yielding higher errors ($46.02\%$ APE for single-extremum queries). Because roughness is defined as the discrepancy between real and idealized perimeters, it relies heavily on fine-grained boundary fidelity, making it inherently sensitive to pixel-level segmentation noise.\\
As introduced in Section~\ref{question_benchmark}, we evaluated performance across two prompt formulations: \textit{clear} (explicit metric naming) and \textit{open} (descriptive, domain-specific terminology). Across all four metrics and task categories, Table~\ref{tab:median_ape_results} indicates nearly identical error profiles between the two query types. This confirms that our system accurately decodes natural, ambiguous biological language into correct analytical operations without sacrificing execution accuracy.
\begin{table}[htbp]
\centering
\caption{Median Absolute Percentage Error (APE) by Metric and Question Type.}
\label{tab:median_ape_results}
\begin{tabular}{lcccc}
\toprule
\multirow{2}{*}{\textbf{Metric}} & \multicolumn{2}{c}{\textbf{Single-Extremum}} & \multicolumn{2}{c}{\textbf{Tertile Mean}} \\
\cmidrule(lr){2-3} \cmidrule(lr){4-5}
& \textbf{Clear Prompt} & \textbf{Open Prompt} & \textbf{Clear Prompt} & \textbf{Open Prompt} \\
\midrule
Area      & 16.77\% & 17.40\% & 38.74\% & 36.77\% \\
Perimeter & 15.36\% & 15.36\% & 22.03\% & 22.24\% \\
Roughness & 46.02\% & 46.02\% & 31.75\% & 33.74\% \\
Roundness &  1.54\% &  1.56\% &  2.48\% &  2.51\% \\
\bottomrule
\end{tabular}
\end{table}

\subsection{Qualitative Comparison against LLM-Based Approaches}
\begin{figure}[htbp]
  \centering
  \begin{subfigure}[b]{0.23\textwidth}
    \includegraphics[width=\textwidth]{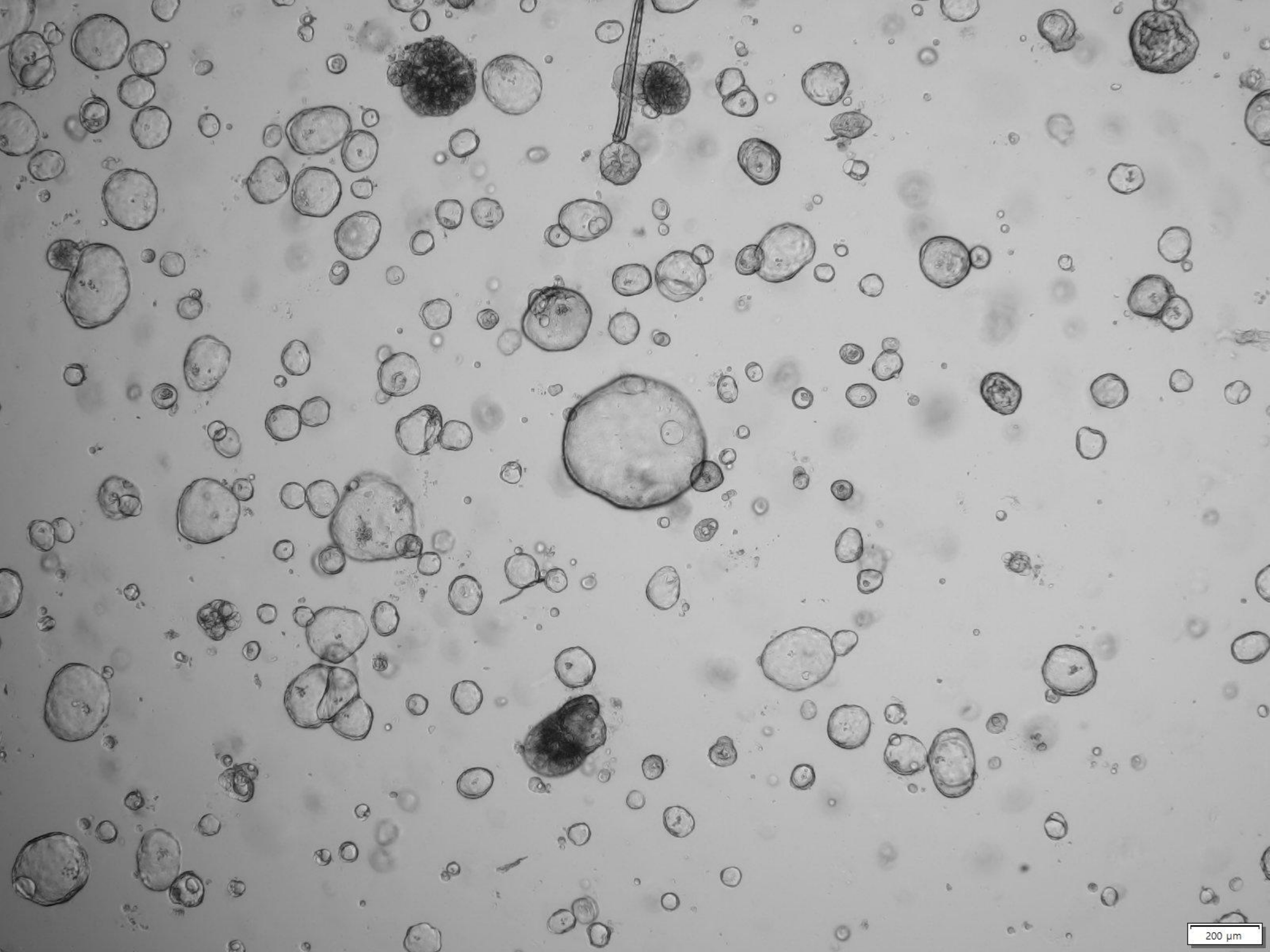}
    \caption{Original}
  \end{subfigure}
  \hfill
  \begin{subfigure}[b]{0.23\textwidth}
    \includegraphics[width=\textwidth]{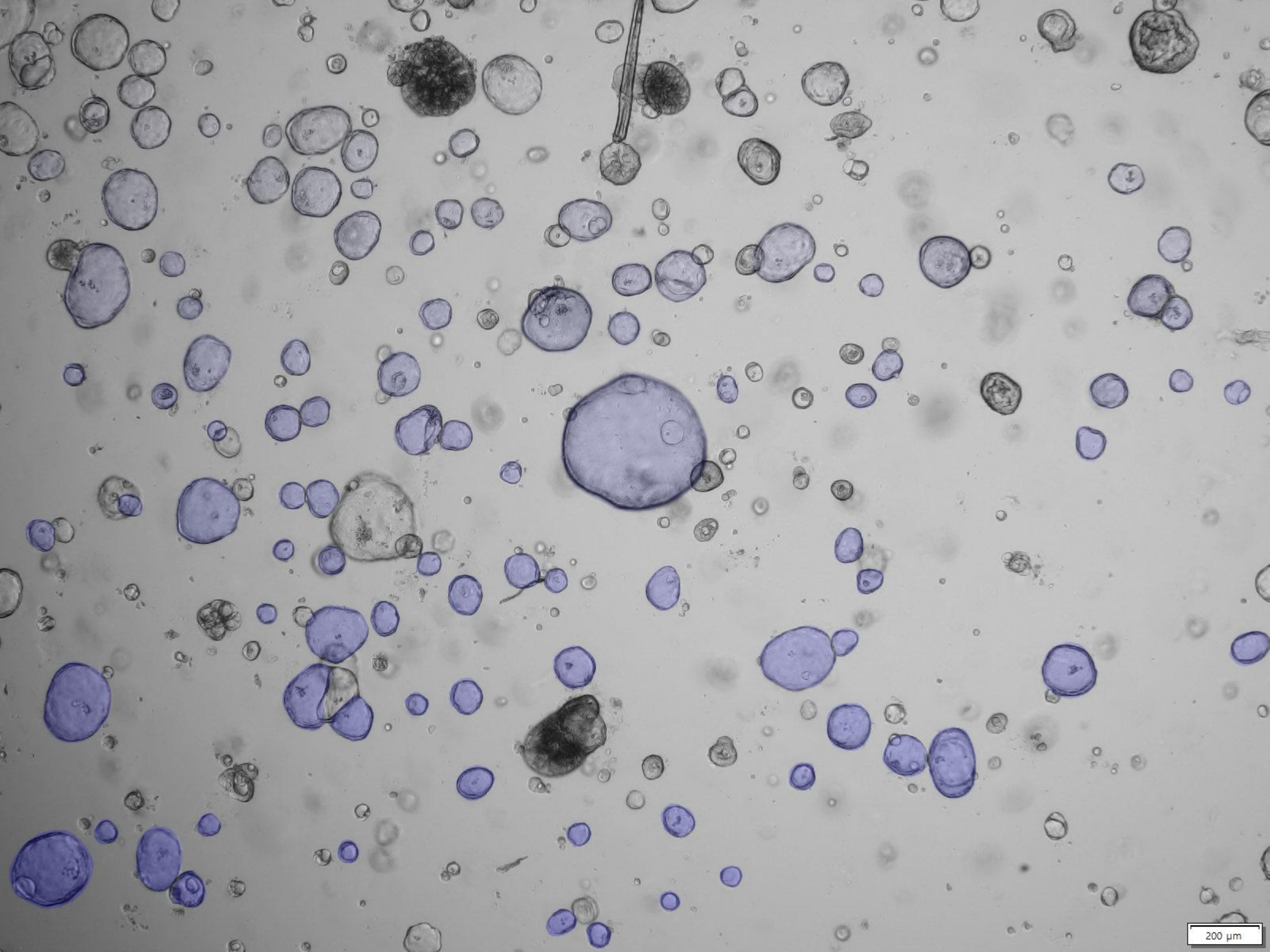}
    \caption{GT}
  \end{subfigure}
  \hfill
  \begin{subfigure}[b]{0.23\textwidth}
    \includegraphics[width=\textwidth]{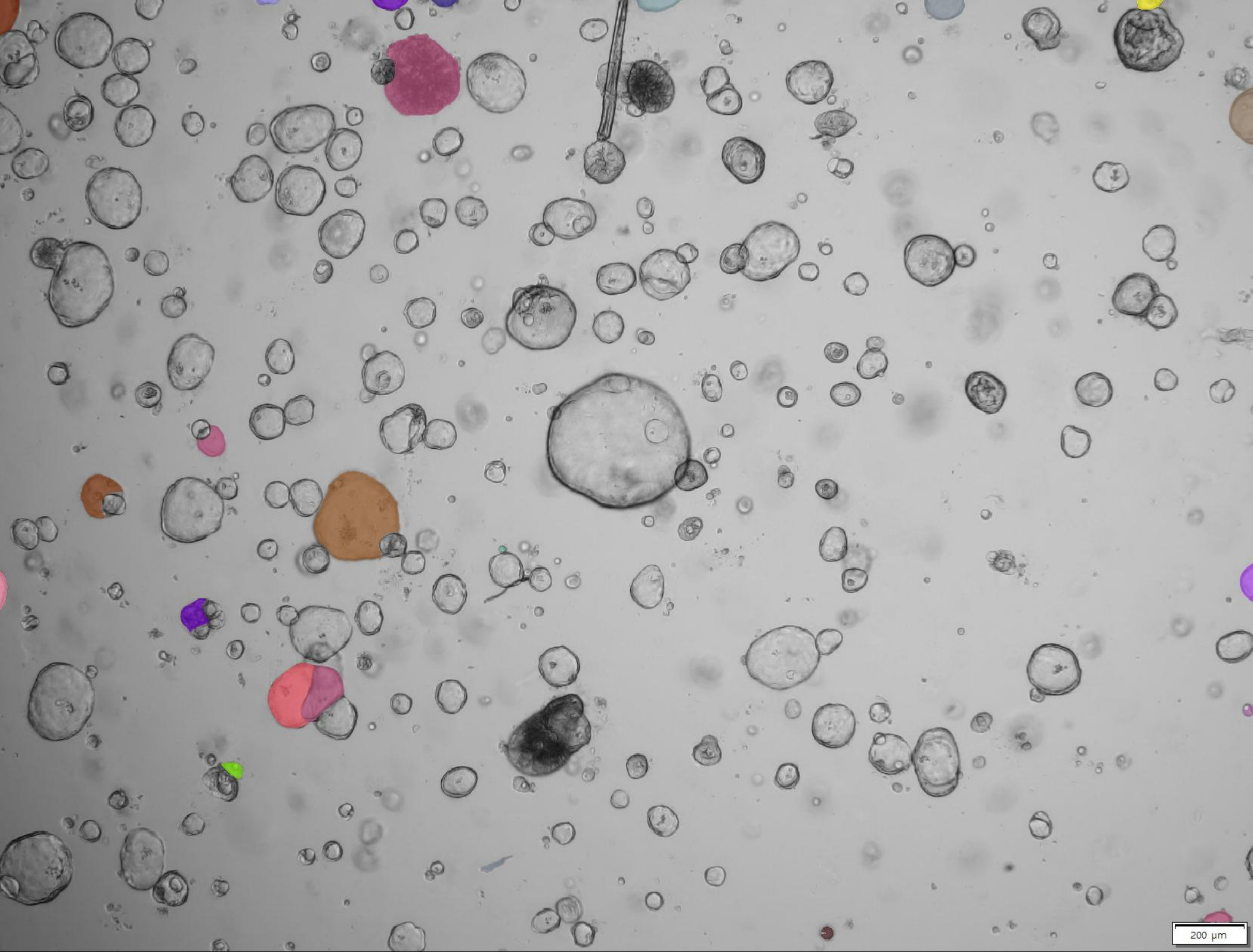}
    \caption{Omega}
  \end{subfigure}
  \hfill
  \begin{subfigure}[b]{0.23\textwidth}
    \includegraphics[width=\textwidth]{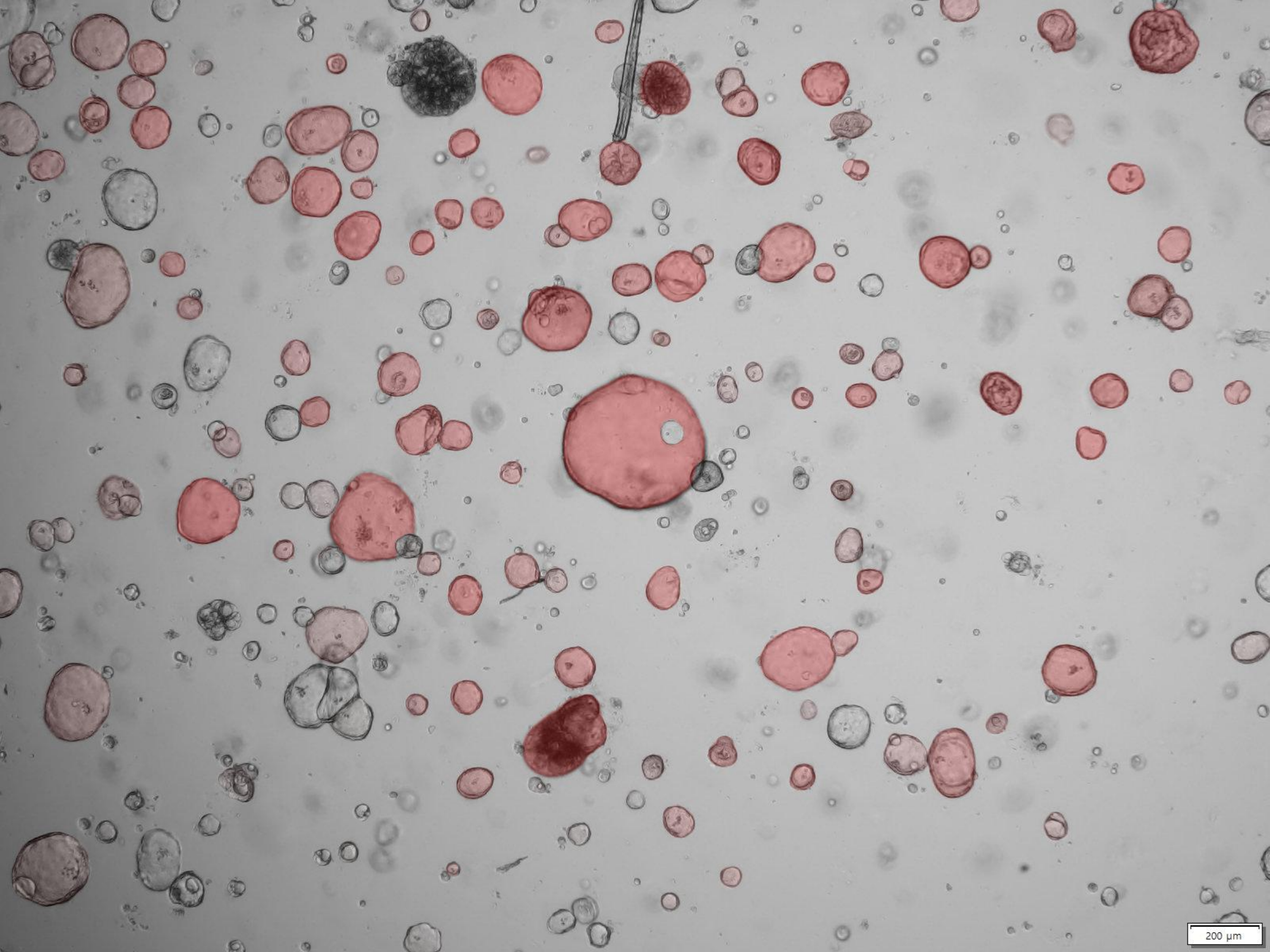}
    \caption{Ours}
  \end{subfigure}
  \caption{Qualitative comparison of organoid segmentation. 
(a) Brightfield image from OrgaExtractor, 
(b) GT segmentation, 
(c) Omega segmentation using StarDist, and 
(d) our hybrid Cellpose--SAM 3 module.
  }
  \label{fig:overlays}
\end{figure}

\begin{figure}[htbp]
  \centering
  \begin{subfigure}[b]{0.3\textwidth}
    \includegraphics[width=\textwidth]{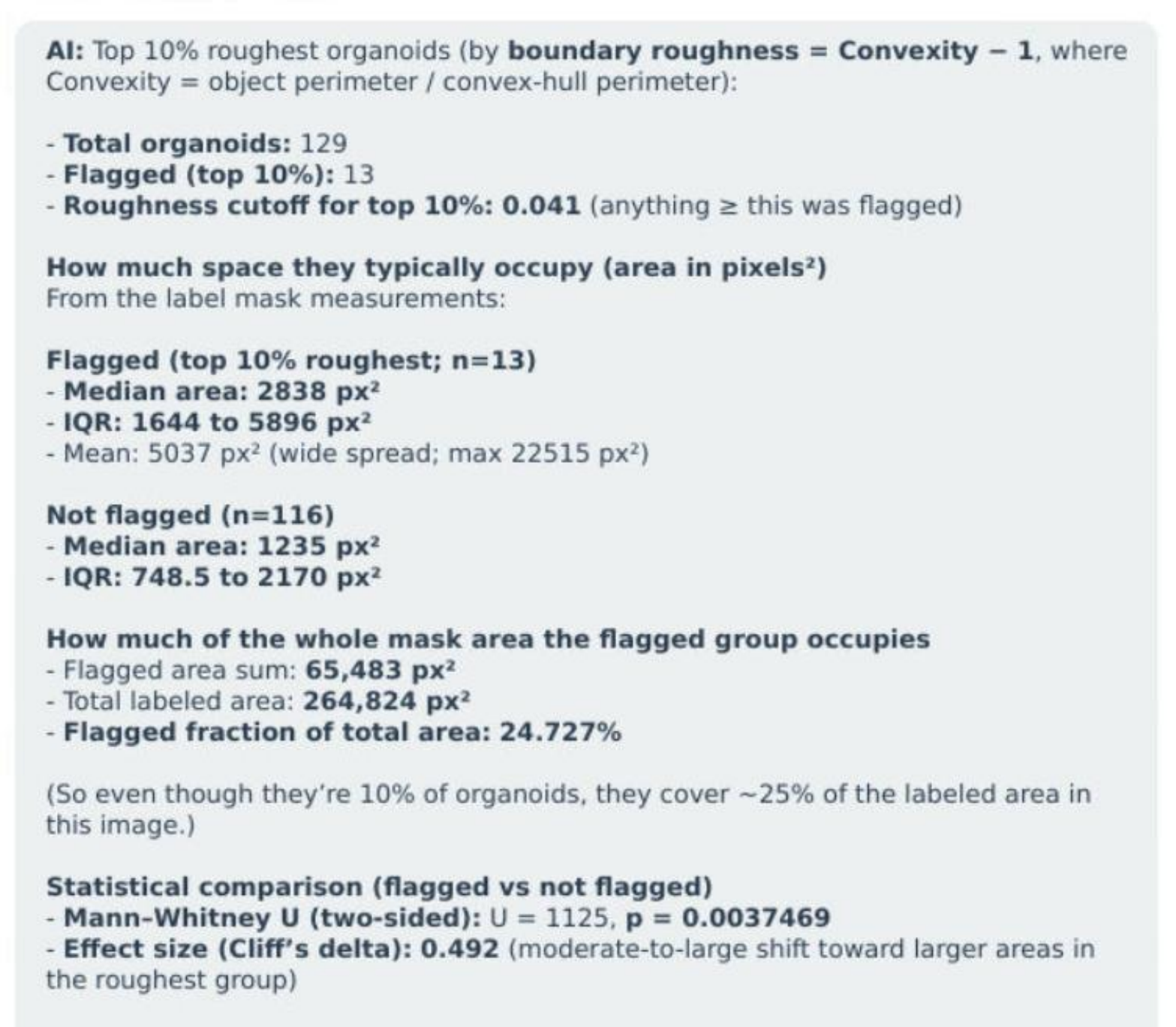}
    \caption{Agentic-J}
  \end{subfigure}
  \hfill
  \begin{subfigure}[b]{0.3\textwidth}
    \includegraphics[width=\textwidth]{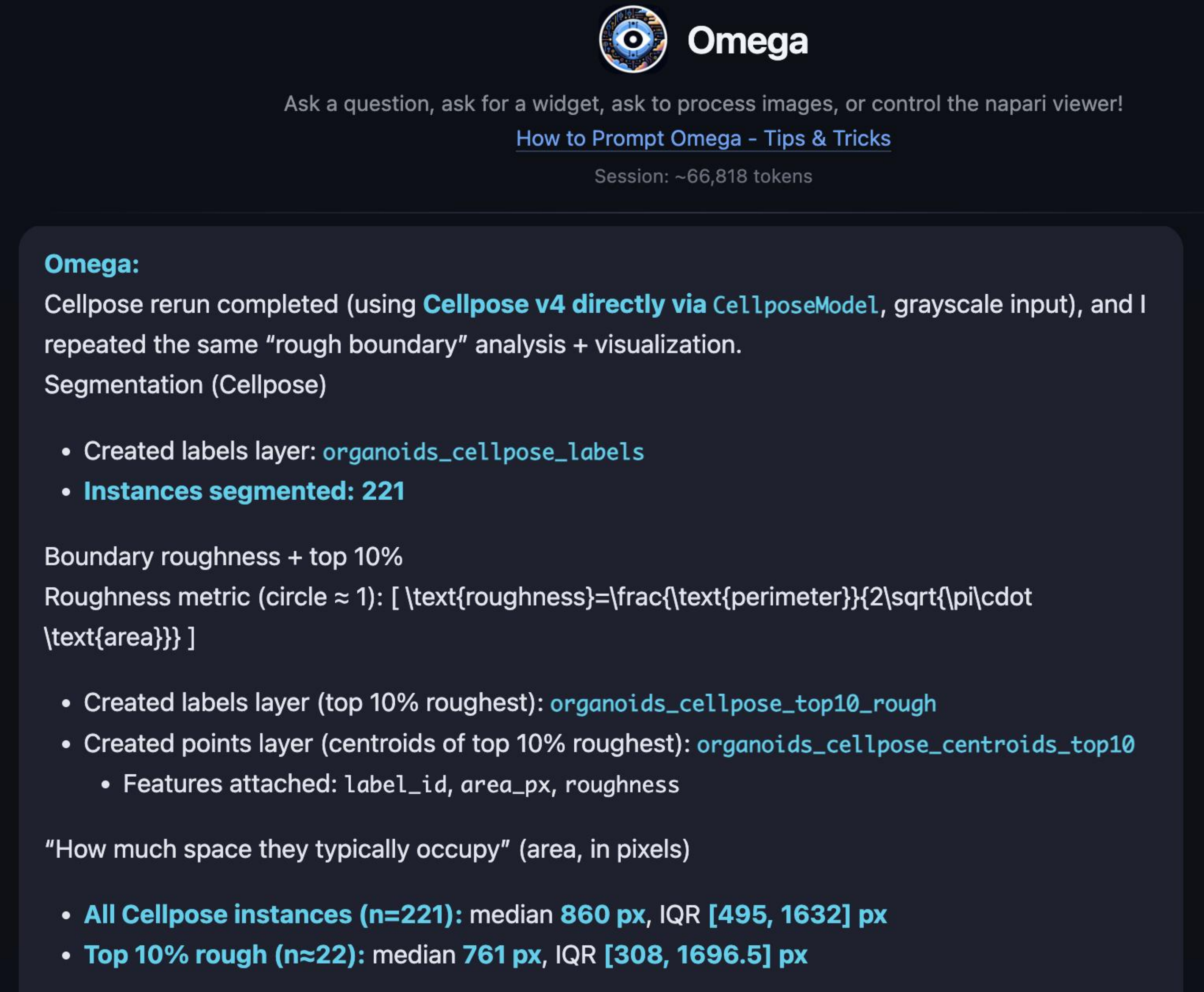}
    \caption{Omega}
  \end{subfigure}
  %\\[0.8em]
  \hfill
  \begin{subfigure}[b]{0.3\textwidth}
    \includegraphics[width=\textwidth]{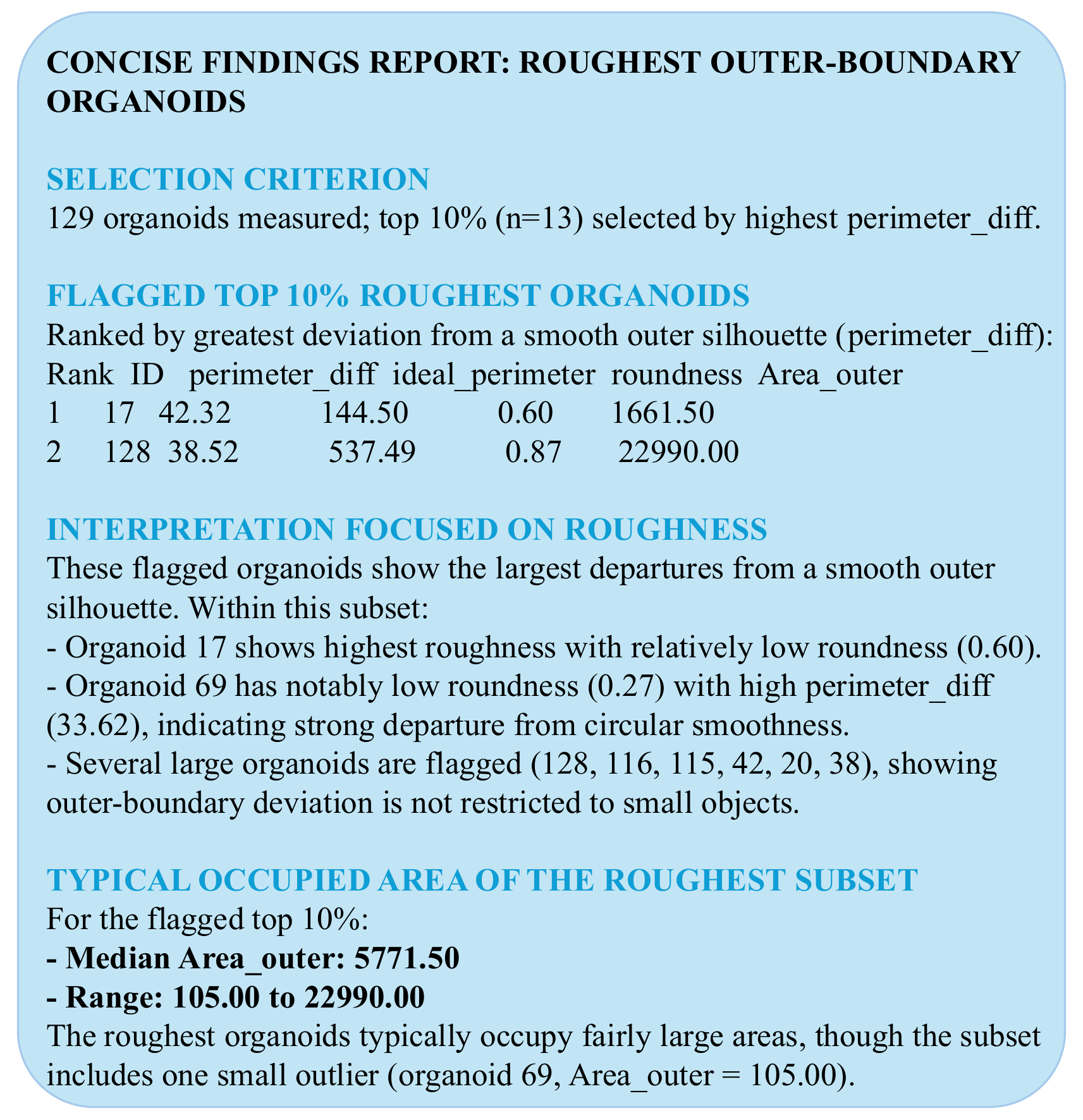}
    \caption{MorphoOrgaAgent}
  \end{subfigure}
  \caption{Comparison of generated reports answering a complex natural-language query requesting boundary roughness filtering (top 10) and spatial area quantification. (a) Agentic-J defaults to standard metrics (e.g., convexity) rather than domain-specific descriptors, yielding an underestimated median area of 1,235 px. (b) Omega produces suboptimal quantitative estimates (median area 761 px) primarily driven by upstream segmentation failures. (c) MorphoOrgaAgent generates a structured, domain-tailored report with exceptional numerical fidelity (median area 5,771.5 px vs. GT 5,749 px; 0.39\% relative error).}
  \label{fig:reports}
\end{figure}

To evaluate the operational workflow and analytical fidelity of our framework, we conduct a qualitative benchmark comparing MorphoOrgaAgent against established bioimage analysis baselines, including Omega~\cite{royer2024omega} and Agentic-J~\cite{ref_agenticj}. The evaluation is performed on a complex, biologically meaningful image-query pair featuring a brightfield microscopy image of colon organoids sourced from the OrgaExtractor dataset~\cite{ref_orgaextractor}. The benchmark query demands cross-metric reasoning and multi-step execution:\textit{``Flag the top 10\% of organoids whose outer boundary strays furthest from a perfectly smooth silhouette, quantify the spatial area they typically occupy, and visualize your findings.''} To answer this question, the system is expected to: (1)~segment all organoids, (2)~compute the difference between real and ideal perimeter, (3)~choose the top 10\% organoids with the highest difference, and (4)~report their median area and visualize the findings.\\
The segmentation results are summarized in Fig.~\ref{fig:overlays}. Although Omega employs StarDist~\cite{stardist1, stardist2, stardist3} for zero-shot segmentation, it exhibits limited generalization on organoid microscopy images. In contrast, our segmentation module yields significantly higher-quality instance masks. This superior zero-shot performance directly stems from our hybrid prompting strategy.
% Row 1: Original image and three overlays
As Agentic-J lacks an integrated deep learning segmentation model, we provided it with the user query alongside our system's predicted mask to generate its final report, as shown in Fig.~\ref{fig:reports}. The median area values reported by Agentic-J, Omega, and MorphoOrgaAgent are $1235$, $761$, and $5771.5$, respectively, against the GT value of $5749$. Notably, MorphoOrgaAgent achieves a minimal relative error of only 0.39\%, demonstrating strong alignment with the ground truth. 
Conversely, Omega yields suboptimal results due to segmentation inaccuracies. While Agentic-J processes the exact same segmentation mask as our model, it defaults to standard metric like convexity to evaluate boundary roughness, missing domain-specific morphological descriptors. It is worth emphasizing that both baseline frameworks are highly capable general-purpose bioimage analysis platforms; however, their generic design limits their ability to capture domain-tailored morphological features compared to MorphoOrgaAgent, which is purpose-built for organoid analysis. As shown in Fig.~\ref{fig:reports}(c), the report generated by MorphoOrgaAgent illustrates the system's step-by-step analysis workflow. First, the framework counts the target organoid population ($N = 129$) and selects the top $10\%$ ($13$ organoids) with the highest boundary roughness ($\text{\texttt{perimeter\_diff}}$). It then uses the computed metrics stored in MorphoOrgaState to rank objects, group subpopulations, and calculate summary statistics.
Because the LLM works strictly as a reasoning engine over pre-computed state data without changing any numbers, our system effectively avoids LLM hallucinations. The resulting report provides clear biological insights: it captures subpopulation variability (distinguishing true boundary irregularity from general low roundness), reports key summary statistics (median area of $5771.5$), and highlights morphological outliers. By combining precise measurements with structured text reports, MorphoOrgaAgent delivers clear and practical results for researchers.
% Row 2: Three reports (2 top, 1 bottom for readability)
\section{Conclusion}
In this work, we present MorphoOrgaAgent, a foundation-model-based multi-agent system that translates complex natural language queries from biologists into structured, machine-executable workflows for organoid morphology analysis. Through a hybrid prompting strategy coupling geometric prompts from Cellpose with domain-tailored text prompts for SAM3, the system enables robust zero-shot organoid segmentation, automated morphological metric computation, visualization, and report synthesis. Quantitative evaluation on our MorphoOrgaVQA benchmark demonstrates that MorphoOrgaAgent effectively interprets both explicit and ambiguous user requests without manual programming. This framework streamlines organoid analysis workflows, reduces expert annotation costs, and paves the way for accessible, high-throughput bioimage intelligence.
\begin{credits}
\subsubsection{\discintname}
The authors have no competing interests to declare that are
relevant to the content of this article.
\end{credits}

\appendix
\section{MorphoOrgaVQA Benchmark}
\label{app:benchmark}

Table~\ref{tab:benchmark_questions} lists the complete set of 16 benchmark
questions, organised by morphological metric, query type (single-extremum
vs.\ tertile mean), and phrasing mode (\emph{clear} vs.\ \emph{open}), together
with the target fields each question is evaluated against. The machine-readable
version (\texttt{benchmark\_questions.json}) and the script that derives the
ground-truth answers from expert-annotated masks are released with our code.

\begin{small}
\begin{longtable}{@{}p{1.1cm} p{7.0cm} p{2.7cm}@{}}
\caption{The complete set of 16 MorphoOrgaVQA benchmark questions.}
\label{tab:benchmark_questions}\\
\toprule
\textbf{Mode} & \textbf{Question} & \textbf{Target fields} \\
\midrule
\endfirsthead
\multicolumn{3}{@{}l}{\textit{(continued from previous page)}}\\
\toprule
\textbf{Mode} & \textbf{Question} & \textbf{Target fields} \\
\midrule
\endhead
\bottomrule
\endlastfoot
\multicolumn{3}{@{}l}{\textit{Area} --- Single-Extremum}\\*
Clear & Identify the specific organoid with the maximum outer area in this field of view, and report both its total cross-sectional surface area in pixels and its geometric center coordinates x and y. & \texttt{Area\_outer}, \texttt{x}, \texttt{y} \\
Open & Locate the most dominant organoid in this image and evaluate how much footprint it occupies in pixel coordinates, alongside its center of mass. & \texttt{Area\_outer}, \texttt{x}, \texttt{y} \\
\addlinespace
\multicolumn{3}{@{}l}{\textit{Area} --- Tertile Mean}\\*
Clear & Perform a stratification of this sample into small, medium, and large tiers based on individual outer area distributions. Provide the mean outer area value calculated for each of the three tiers. & \texttt{Area\_outer} \\
Open & Quantify the size heterogeneity of this culture by partitioning all organoids into three tiers. What is the mean surface area for each of the three tiers---small, medium, and large? & \texttt{Area\_outer} \\
\addlinespace
\multicolumn{3}{@{}l}{\textit{Perimeter} --- Single-Extremum}\\*
Clear & Identify the specific organoid with the maximum outer perimeter in this field of view, and report its total boundary length in pixels. & \texttt{Perimeter\_outer} \\
Open & Identify the single organoid that possesses the longest external boundary line and report its total boundary length in pixels. & \texttt{Perimeter\_outer} \\
\addlinespace
\multicolumn{3}{@{}l}{\textit{Perimeter} --- Tertile Mean}\\*
Clear & Stratify this entire population into three distinct tiers using outer perimeter limits. Compute the mean outer perimeter value for each of the three tiers. & \texttt{Perimeter\_outer} \\
Open & If we split this entire population into three tier levels based on the length of their edges, what is the mean perimeter value for each of the three tiers? & \texttt{Perimeter\_outer} \\
\addlinespace
\multicolumn{3}{@{}l}{\textit{Roundness} --- Single-Extremum}\\*
Clear & Find the single organoid that records the absolute maximum score for roundness in this imaging frame, and report its circularity index value & \texttt{roundness} \\
Open & Scan the image and pinpoint the single most symmetrical, spherical organoid in this batch. What is its exact circularity score? & \texttt{roundness} \\
\addlinespace
\multicolumn{3}{@{}l}{\textit{Roundness} --- Tertile Mean}\\*
Clear & Stratify the organoid population into three circularity tiers based on its roundness score. Provide the average roundness value for each of the three tiers. & \texttt{roundness} \\
Open & Perform a quality control triage by dividing all organoids into three structural shape classes---spherical, intermediate, and irregular. Give me the average circularity for each of the three classes. & \texttt{roundness} \\
\addlinespace
\multicolumn{3}{@{}l}{\textit{Roughness} --- Single-Extremum}\\*
Clear & Locate the organoid showing the maximum baseline perimeter difference relative to its calculated ideal perimeter. Output its raw boundary excess value in pixels. & \texttt{perimeter\_diff} \\
Open & Target the anomalous organoid that deviates the most from a smooth track---possessing the highest morphological roughness---and report its exact boundary line excess in pixels. & \texttt{perimeter\_diff} \\
\addlinespace
\multicolumn{3}{@{}l}{\textit{Roughness} --- Tertile Mean}\\*
Clear & Bin all organoids into three classes according to their perimeter difference variance. Provide the mean perimeter difference value for each of the three classes---low, medium, and high deviation groups. & \texttt{perimeter\_diff} \\
Open & Classify this population into three tiers ranging from uniform to highly folded margins. What is the average roughness metric for each of the three tiers? & \texttt{perimeter\_diff} \\
\addlinespace
\end{longtable}
\end{small}

\section{System Prompts of the LLM Agents}
\label{app:prompts}

For full reproducibility, we list below the verbatim system prompts of the two
LLM-driven subagents in MorphoOrgaAgent. The TaskUnderstandingAgent
(Appendix~\ref{app:prompt_tua}) is constrained to emit a strict
\texttt{AnalysisIntent} JSON object, which restricts its output to the
predefined metric and visualization pools and makes the produced analysis plan
machine-checkable. The ReportAgent (Appendix~\ref{app:prompt_ra}) is
deliberately kept minimal and is explicitly forbidden from recomputing or
overriding any numeric value, which is the mechanism by which the framework
avoids numerical hallucination.

\subsection{TaskUnderstandingAgent}
\label{app:prompt_tua}
\begin{lstlisting}[basicstyle=\ttfamily\scriptsize, breaklines=true,
  columns=fullflexible, keepspaces=true, frame=single, captionpos=t,
  caption={System prompt of the TaskUnderstandingAgent.}, label={lst:tua}]
You are TaskUnderstandingAgent for organoid microscopy image analysis.

Your job is to convert the user's natural-language request into a strict AnalysisIntent JSON object for downstream execution modules.
Segmentation is always required by the pipeline, so do not decide whether segmentation is needed.
Only decide which quantitative metrics and which visualization categories are needed.

=========================================
METRIC POOL
=========================================
required_metrics must contain only metric names from this pool:
- organoid_idx: The sequential category ID allocated to each distinct organoid.
- Area_outer: Total surface cross-sectional area enclosed by the outermost boundary in pixels.
- Perimeter_outer: The curve length of the organoid's external boundary silhouette in pixels.
- areas_inner: Cumulative surface area of all inner cavities or lumens nested inside the organoid body in pixels.
- perimeters_inner: Cumulative boundary contour length of all inner cavities or hollow lumens in pixels.
- x: Spatial horizontal coordinate of the geometric centroid (center of mass).
- y: Spatial vertical coordinate of the geometric centroid (center of mass).
- is_border: A boolean flag tracking if the organoid touches the image frame edge (used to filter out incomplete/truncated shapes).
- ideal_perimeter: Theoretical perimeter length calculated if the organoid was a mathematically perfect circle with the same outer area.
- perimeter_diff: The morphological rugosity indicator (Actual external perimeter minus the ideal perimeter).
- roundness: Isoperimetric circularity value ranging from 0.0 (irregular/spiky) to 1.0 (perfect circle).
- lumen_ratio: Spatial ratio of internal hollow cavity space relative to the outer dimension (areas_inner / Area_outer), used to track luminal differentiation events.
- area_outer_microns: Physical surface dimensions converted into squared microns based on spatial calibration.
- area_outer_mm: Tangible physical surface area translated into squared millimeters for standard biological reasoning reports.
- radius_from_peri: Derived organoid equivalent radius reversed analytically from the outer perimeter parameter.
- area_from_peri: Comparative reference surface area reversed analytically from the outer perimeter parameter.

=========================================
CRITICAL DECISION LAWS FOR VISUALIZATIONS
=========================================
required_visualizations MUST contain ONLY values from this pool, mapped strictly by these logic laws:
- pixel_overlays: Pixel-level overlay visualizations on the raw microscopy image, including colored instance mask overlays and bounding-box overlays. Use when the user wants to see segmentation quality, locate objects, count or identify individual organoids, inspect boundaries, or verify which pixels belong to each organoid.
- metric_plots: Metric-derived visualizations, including metric heatmap overlays on the image, centroid spatial mapping, and metric distribution plots. Use when the user asks about area, perimeter, roundness, lumen ratio, size distribution, morphology distribution, spatial layout, centroid positions, clustering, outliers, heterogeneity, or any metric value mapped back to objects.
- correlation_heatmap: Higher-level association visualization showing correlations among multiple numeric organoid metrics. Use when the user asks about relationships, associations, correlations, trade-offs, covariance, feature interactions, or how metrics vary together.
- none: Use ONLY when the user explicitly requests numeric/text output without any visualization, or when no visualization is implied by the request.

If more than one visualization category is useful, include all useful categories.
Do not return old concrete chart names such as mask_overlay, area_histogram, roundness_histogram, scatter_area_roundness, spatial_map, time_series_plot, or qc_summary_plot.

=========================================
REPORT DISPATCHING MANDATE
=========================================
You are the definitive dispatcher for the final ReportAgent. The report_instruction field is a compressed, highly directive handoff prompt for that downstream agent.

CRITICAL CO-REFERENCE RULE: This field MUST NEVER BE EMPTY and MUST BE DYNAMICALLY TAILORED.
You must explicitly read your own chosen tokens in `required_metrics` and `required_visualizations`, and mention them inside the `report_instruction` narrative.

- If the user has specific formatting guidelines: extract them, and weave them together with your chosen metrics and plots. (e.g., "Analyze the calculated roundness and Area_outer values, look at the generated metric_plots, and explain the morphological variance in concise Chinese bullet points.")
- If the user has NO specific formatting guidelines: actively convert their request into an explicit operational directive that lists what data they will receive. (e.g., "Draft a professional analysis report detailing the total count and layout from the calculated x and y coordinates and the planned pixel_overlays visualization.")

=====================================
OTHER GENERAL CONSTRAINTS
=====================================
Return strict JSON only. Do not include Markdown, ```json, comments, or explanatory text.
If the request is vague, still return JSON and lower confidence.
Do not ask the user any question. Make the best possible interpretation.
target_objects should include "organoid" by default.
confidence must be between 0 and 1.
\end{lstlisting}

\subsection{ReportAgent}
\label{app:prompt_ra}

\begin{lstlisting}[basicstyle=\ttfamily\scriptsize, breaklines=true,
  columns=fullflexible, keepspaces=true, frame=single, captionpos=t,
  caption={System prompt of the ReportAgent.}, label={lst:ra}]
You are ReportAgent for organoid microscopy image analysis.

Your only job is to write a markdown scientific report that directly answers the provided report_instruction using the provided OrganoidState report context.

Mandatory constraints:
- Only use evidence from OrganoidState.
- Do not invent measurements, object counts, file paths, images, masks, visualizations, or conclusions.
- Do not modify, reinterpret, recalculate, or override object count or metric values.
- If evidence is missing, explicitly state that it is unavailable.
- Mention limitations and uncertainty.
- Answer the report_instruction directly.
- Do not claim that segmentation, quantification, plotting, or any other analysis was rerun.
- Use quantitative values exactly as provided in the context.
- If visual evidence is available only as paths or array summaries, describe it as available evidence without pretending to visually inspect pixels.

Return markdown only.
\end{lstlisting}

\end{document}